\documentstyle[aps]{revtex}
\begin{document}
\voffset 0.8truecm
\title{ 
Quantify entanglement by concurrence hierarchy
}
\author{
Heng Fan, Keiji Matsumoto, Hiroshi Imai
}
\address{
Quantum computing and information project,
ERATO, \\
Japan Science and Technology Corporation,\\
Daini Hongo White Bldg.201, Hongo 5-28-3, Bunkyo-ku, Tokyo 133-0033, Japan.
}
\maketitle
                                                        
\begin{abstract}
We define the concurrence hierarchy as $d-1$ independent invariants
under local unitary transformations in d-level quantum system.
The first one is the
original concurrence defined by
Wootters et al \cite{W,HW} in
2-level quantum system and generalized to
d-level pure quantum states case.
We propose to use this concurrence hierarchy as measurement of
entanglement. This measurement does not increase under local
quantum operations and classical communication.
\end{abstract}
       

\section{Introduction}
Entanglement plays a central role in quantum computation and
quantum information\cite{NC}. One of the main goals of theory
of entanglement is to develope measures of entanglement.
Several measures of entanglement are proposed and studied
according to different aims,
including entanglement of formation,
entanglement of distillation, entanglement cost etc.\cite{BDSW,BBPS}.

Perhaps one of the most widely accepted measures of entanglement
is entanglement of formation $E_f$ which provides a very
good measurement of entanglement asymptotically.
For a pure bipartite quantum state
$\rho =|\Phi \rangle \langle \Phi |$ shared by
A and B,
entanglement of formation
is defeined by von Neumann entropy of reduced density matrix
$E_f(\rho )=-Tr\rho _A\log \rho _A$, where $\rho _A=Tr_B\rho $. 
For mixed state, the entanglement of formation takes the form
\begin{eqnarray}
E_f(\rho )={\rm inf}\sum _j{p_j}E_f(\Phi _j),
\end{eqnarray}
where the infimum is taken over all pure-state decompositions
of $\rho =\sum _jp_j|\Phi _j\rangle \langle \Phi _j|$.
For mixed state,
this definition is operational difficult because it
requires finding the minimum average entanglement
over all possible pure-state decompositions of the given mixed
state.
In d-dimension, the explicit expression of entanglement of formation
is only found for several special types of
mixed state, for example, the isotropic states \cite{TV}
and Werner states \cite{VW}.
However, the explicit formulas have been found for
the 2-level quantum system by Wootters et al\cite{W,HW}.
Here we briefly introduce the results by Wootters et al.
The entanglement of formation of an arbitrary state $\rho $
is related to a quantity called {\it concurrence} $C(\rho )$
by a function
\begin{eqnarray}
E_f(\rho )=\epsilon (C(\rho ))
=h\left( \frac {1+\sqrt{1-C^2(\rho )}}{2}\right),
\end{eqnarray}
where $h(x)=-x\log x-(1-x)\log (1-x)$ is the binary
entropy function. The entanglement of formation is
monotonically increasing with respect to the
increasing concurrence. The concurrence is defined
by an almost magic formula,
\begin{eqnarray}
C(\rho )=max\{0, \lambda _1-\lambda _2-\lambda _3-\lambda _4\}
\label{magic}
\end{eqnarray}
where the $\lambda _i$'s are the square root of the
eigenvalues of $\rho \tilde {\rho }$ in descending order.
And $\tilde {\rho }=(\sigma _y\otimes \sigma _y)\rho ^*
(\sigma _y\otimes \sigma _y)$,
where $\sigma _y$ is the Pauli matrix. For pure state
$|\Phi \rangle =\alpha _{00}|00\rangle +\alpha _{01}|01\rangle
+\alpha _{10}|10\rangle +\alpha _{11}|11\rangle $,
the concurrence takes the form
\begin{eqnarray}
C(\Phi )=|\langle \Phi |\sigma _y\otimes \sigma _y|\Phi ^*\rangle |
=2|\alpha _{00}\alpha _{11}-\alpha _{01}\alpha _{10}|.
\end{eqnarray}
Because of the relation between concurrence and entanglement
of formation, we can use directly the concurrence as the
measure of entanglemnet.

One important aim in formulating the measures of entanglement
is to find whether a bipartite state is separable or not
because the entanglement state has some useful
applications, for example, teleportation
\cite{BBCJPW} quantum cryptography
by using EPR pairs\cite{E}.
In 2-level quantum system, the Peres-Horodeckis\cite{P,H} criterion
is a convenient method. And the concurrence provide another
method. If the concurrence is zero, the quantum state is separable,
otherwise it is entangled. For general mixed state in d-dimension,
we need yet to find an operational method to distinguish
separability and entanglement. 

For pure state in d-dimension,
the measure of entanglement is largely solved
by entanglement of formation. We can use it to distinguish
whether a pure state is separable or not and to find
the amount of entanglement.
However, to complete characterizing the entanglement, one
quantity seems not enough.
A simple example is\cite{HHH}:
\begin{eqnarray}
|\psi \rangle &=&1/\sqrt{2}(|00\rangle +|11\rangle ),
\nonumber \\
|\phi \rangle &=&\sqrt{x}/\sqrt{2}(|00\rangle +|11\rangle )+
\sqrt{1-x}|22\rangle,
\label{example1}
\end{eqnarray}
When $x\approx 0.2271$ is a root of equation $x^x[2(1-x)]^{1-x}=1$,
the entanglement of formation equal to 1 for both $|\psi \rangle $ and
$|\phi \rangle $.
However, they can not be transformed to each other by
local operations and classical communication (LOCC).

Because concurrence provide a measure of entanglement
in 2-level system, it is worth generalizing concurrence to
higher dimension. There are several proposals for the case of
pure states\cite{U,RBCHM,AF,HHH,ASST,ZZF}. Uhlmann generalized the
concurrence by considering arbitrary conjugations acting
on arbitrary Hilbert spaces\cite{U}. Rungta et al generalized
the spin flip operator $\sigma _y$ to a universal inverter
$S_d$ defined as $S_d(\rho )=1-\rho $, so the pure state
concurrence in any dimension takes the form
\begin{eqnarray}
C'(\Phi )&=&\sqrt {\langle \Phi |S_{d_1}\otimes S_{d_2}
(|\Phi \rangle \langle \Phi |)|\Phi \rangle }
\nonumber \\
&=&\sqrt{2[1-Tr(\rho _A^2)]}.
\end{eqnarray}
There is a simple relation between these
two generalizations pointed out by Wootters\cite{W1}.
Another generalization proposed by Albeverio and Fei\cite{AF} by
using an invariants under local unitary transformations
turns out to be the same as that of Rungta et al up to a whole factor.
They define the concurrence as
\begin{eqnarray}
C(\Phi )=\sqrt{\frac {d}{d-1}[1-Tr(\rho _A^2)]}.
\label{2concurrence}
\end{eqnarray}
Let's analyze the example (\ref{example1}) again by the
generalized concurrence. When $x=1/3$ is a root of
equation $(3x-1)(x-1)=0$, the concurrence of
$|\psi \rangle $ and $|\phi \rangle $ are equal.
But still $|\psi \rangle $ and $|\phi \rangle $
can not be transformed to each other by LOCC.

As already noticed and conjectured by many researchers, one
quantity perhaps is not enough to measure all aspects of
entanglement\cite{V1,JP,JP1,H0,V2,VDM,ZB,SZK}, see
\cite{VN} for a review, and the geometric
properties of entanglement was investigated
in \cite{KZ}. As the question of separability,
Peres-Horodeckis \cite{P,H}
criterion is enough for bipartite 2-level quantum system.
For higher dimension,
if we want to find whether a bipartite state is entangled,
besides partial transposition operation proposed by Peres\cite{P},
we
need to find other positive but not completely positive maps.
Presently, how to find whether a bipartite state in
$C^{d_1}\times C^{d_2}$ is entangled
is still an open problem.

\section{Definition of concurrence hierarchy}
In this paper, we propose to use the concurrence hierarchy
to quantify the entanglement for d-dimension.
We restrict ourself to $C^d\otimes C^d$ bipartite pure
state. 
A general bipartite pure state in $C^d\otimes C^d$ can
be written as
\begin{eqnarray}
|\Phi \rangle =\sum _{i,j=0}^{d-1}\alpha _{ij}|ij\rangle ,
\label{phi}
\end{eqnarray}
with normalization $\sum _{ij}\alpha _{ij}\alpha ^*_{ij}=1$.
We define a matrix $\Lambda $ with entries
$\Lambda _{ij}=\alpha _{ij}$.
The reduced density matrix can be denoted as
$\rho _A=Tr_B\rho =\Lambda \Lambda ^{\dagger }.$
Under a local unitary transformation
$U\otimes V$, the matrix $\Lambda $ is changed to
$\Lambda \rightarrow U^t\Lambda V$, where the superindex $t$ represents
transposition. And the redeced density operator thus
is transformed to
\begin{eqnarray}
\rho _A\rightarrow (U^t\Lambda V)(V^{\dagger }
\Lambda ^{\dagger } U^{t\dagger })
=U^t\Lambda \Lambda ^{\dagger }U^{t\dagger }.
\end{eqnarray}
In 2-dimension, it
was point out by Linden and Popescu \cite{LP},
there is one no-trivial invariant under local unitary
transformations $I=Tr(\Lambda \Lambda ^{\dagger })^2$.
In general d-dimension, it was pointed out by Albeverio and Fei that
there are $d-1$ independent invariants under local
unitary transformations $I_{k}=Tr(\Lambda \Lambda ^{\dagger })^{k+1}$.
When $k=0$, it is just the normalization equation
$I_0=\sum _{ij}\alpha _{ij}\alpha _{ij}^*=1$.
For $k=1, \cdots, d-1$, $I_k$ are $d-1$ independent invariants
under local unitary transformations. Then they generalize
the concurrence as the formula (\ref{2concurrence}) and
one relation be calculated as
\begin{eqnarray}
1-Tr\rho _A^2=I_0-I_1=\frac {1}{2}\sum _{i,j,k,m}^d
|\alpha _{ik}\alpha _{jm}-\alpha _{im}\alpha _{jk}|^2.
\label{intuitive}
\end{eqnarray}
When $C(\Phi )=0$, it is separable; when $C(\Phi )\not= 0$, it is
entangled; when $C(\Phi )=1$, it is maximally entangled state.
For a pure state $|\Phi \rangle $ as in (\ref{phi}), when
$\alpha _{ik}\alpha _{jm}=\alpha _{im}\alpha _{jk}$ for all
$i,j,k,m$, it can be written as a product form and thus
separable. It is a rather intuitive idea to use quantity
(\ref{intuitive}) as the measure of entanglement.
And all proposals of generalization of concurrence
actually lead to this result. And also
when $C(\Phi )\not= 0$, state $|\Phi \rangle $ is entangled.
However, our opinion is that this quantity is necessary but
not enough.
In quantifying the entanglement, 
the entanglement is dealed independently by restricted to every
2-level system.
For example, suppose $|\Phi '\rangle $ takes
the form
\begin{eqnarray}
|\Phi '\rangle =\alpha _{00}|00\rangle +\alpha _{11}|11\rangle
+\alpha _{22}|22\rangle .
\label{example2}
\end{eqnarray}
Actually we can always change
a pure state $|\Phi \rangle $ to this form by Schmidt decomposition.
The states $\alpha _{00}|00\rangle +\alpha _{11}|11\rangle $,
$\alpha _{00}|00\rangle +\alpha _{22}|22\rangle $ and
$\alpha _{11}|11\rangle +\alpha _{22}|22\rangle $
are considered independently in (\ref{intuitive}) and
the entanglement in every 2-level system is sumed together
$C(\Phi )=|\alpha _{00}\alpha _{11}|^2+|\alpha _{00}\alpha _{22}|^2
+|\alpha _{11}\alpha _{22}|^2$. As already pointed out previously,
when $x=1/3$, the concurrence of $|\psi \rangle $ and
$|\phi \rangle $ in (\ref{example1})
are equal but they can not be transformed to
each other by LOCC. Our idea here is that besides the concurrence
in the form (\ref{intuitive}), we should also quantify it
by other quantities. For example, the state
$|\Phi '\rangle $ in (\ref{example2}), we can quantify the
entanglement by
\begin{eqnarray}
C_3(\Phi ') =|\alpha _{00}\alpha _{11}\alpha _{22}|^2,
\label{3concurrence}
\end{eqnarray}
up to a normalized factor.
In this quantity we just consider the entanglement in all 3 levels.
Apparently, $C_3(\Phi ')=0$ does not mean
the state $|\Phi '\rangle $ is separable.
So both this quantity and (\ref{intuitive}) are necessary
in quantifying the entanglement in 3-level quantum system.
We call these two quantities as {\it concurrence hierarchy} for
3-level system.
The example (\ref{example1}) thus can be distinguished as follows.
If you let both $C(\psi )=C(\phi )$ and 
$C_3(\psi )=C_3(\phi )$, we can find just one solution
$x=1$, i.e. $|\psi \rangle =|\phi \rangle $.
In case $x=1/3$, though the two level concurrences defined in
(\ref{2concurrence}) for $|\psi \rangle $ and $|\phi \rangle $ are equal,
their 3-level concurrences are different, $C_3(\psi )=0$
while $C_3(\phi )=1/54$. The
structure of their concurrence hierarchy is different. So, they can not
be transformed to each other by LOCC.

Next, we give our precise definition of concurrence hierarchy.
Suppose a bipartite pure state (\ref{phi}) shared by A and B,
$\lambda _{\Phi }=\{ \lambda _0^{\downarrow },
\cdots ,\lambda _{d-1}^{\downarrow }\} $ denotes the vector of
eigenvalues of the reduced density operator $\rho _A=Tr_B(|\Phi \rangle
\langle \Phi |)$ in decreasing order.
In other words $\lambda _j^{\downarrow},j=0,\cdots ,d-1$ are square of
singular values of matrix $\Lambda $.

Definition: {\it The concurrence hierarchy of the state $|\Phi \rangle $
is defined as
\begin{eqnarray}
C_k(\Phi )&=&\sum _{0\le i_0<i_1<\cdots <i_k\le (d-1)}
\lambda _{i_0}^{\downarrow }\lambda _{i_1}^{\downarrow }
\cdots \lambda _{i_k}^{\downarrow },
\nonumber \\
&&~k=1, 2, \cdots, d-1.
\label{hierarchy}
\end{eqnarray}
We propose to use this concurrence hierarchy to quantify the
entanglement of the state $|\Phi \rangle $.}

The first level concurrence is trivial since it is just
the normalization condition
$C_1(\Phi )=\sum _{i=0}^{d-1}\lambda _i^{\downarrow }=1$.
The two level concurrence is the d-dimension
generalization of concurrence proposed by
Rungta et al \cite{RBCHM} and Albeverio et al\cite{AF} up to
a whole factor. In 2-dimension, there are just one
non-trivial concurrence which is the original concurrence
proposed by Wootters et al\cite{W,HW}. In d-dimension, the
concurrence hierarchy consists of $d-1$ independent non-trivial
concurrences. The result of 3 level concurrence in
3-dimension is already presented in (\ref{3concurrence}).
This concurrence hierarchy is invariant under local unitary
transformations and can be represented in terms of 
invariants $I_{k}=Tr(\Lambda \Lambda ^{\dagger })^{k+1}$ \cite{AF}.
It should be noted that a similar idea as this paper
was also proposed by Sinolecka, Zyczkowski and Kus \cite{SZK}.
We give an example to show one relation for 3 level concurrence
of state $|\Phi \rangle $ in (\ref{phi}),
\begin{eqnarray}
C_3(\Phi )&=&\sum _{0\le i_0<i_1<i_2\le (d-1)}
\lambda _{i_0}^{\downarrow }\lambda _{i_1}^{\downarrow }
\lambda _{i_2}^{\downarrow }
\nonumber \\
&=&1+2I_2-3I_1
\nonumber \\
&=&{1\over 6}\sum _{ijklmr}|
\alpha _{ij}\alpha _{kl}\alpha _{mr}
+\alpha _{kj}\alpha _{ml}\alpha _{ir}
+\alpha _{mj}\alpha _{il}\alpha _{kr}
\nonumber \\
&&-\alpha _{mj}\alpha _{kl}\alpha _{ir}
-\alpha _{il}\alpha _{kj}\alpha _{mr}
-\alpha _{ij}\alpha _{ml}\alpha _{kr}|^2,
\label{3levelh}
\end{eqnarray}
where terms inside $|\cdot |$ correspond to
determinants of the $3\times 3$ submatrix of
$\Lambda $ with row indices $i,k,m$
and column indices $j,l,r$.
When $|\Phi \rangle $ is separable, all concurrences in the
hierarchy are zeros except the trivial one.
If the Schmidt number (rank) of $\rho _A$ for state
$|\Phi \rangle $ in (\ref{phi}) is $k, 1\le k\le d$, all
higher level concurrences $C_j(\Phi )=0, j>k$.
This is simple because all eigenvalues of $\rho _A$ are non-negative.

\section{A simple method to calculate the concurrence hierarchy and
entanglement can be quantified by concurrence hierarchy}

The concurrence hierarchy can be calculated
by its definition (\ref{hierarchy}).
The 2,3-level concurrences can be calculated
directly by relations (\ref{intuitive},\ref{3levelh}).
Here we show all concurrences in the hierarchy can
be calculated similarly. According to some
results in linear algebra, see for example Ref.\cite{B},
the concurrence hierarchy $C_k(\Phi )$ equal to the
sums of the $k$-by-$k$ principal minors of reduced
density operator $\Lambda \Lambda ^{\dagger }$.
And it is known that these quantities are
invariant under unitary transformations          
$U\Lambda \Lambda ^{\dagger }U^{\dagger }$.
This leads straightforward to the result that for a bipartite
pure state (\ref{phi}), the concurrence hierarchy
$C_k(\Phi )$ are invariant under local unitary
transformations. For convenience, we adopt the same
notations as that of Ref.\cite{B}.
Let $\beta ,\gamma \subseteq \{0,...,d-1\} $ be
index sets, each of cardinality $k$, $k=1, \cdots, d$.
According to Cauchy-Binet formula, we have
the following relations:
\begin{eqnarray}
C_k(\Phi )&=&\sum _{\beta }\det \rho _A(\beta ,\beta )
\nonumber \\
&=&\sum _{\beta }\sum _{\gamma }\det \Lambda (\beta ,\gamma )
\det \Lambda ^{\dagger }(\gamma ,\beta )
\nonumber \\
&=&\sum _{\beta }\sum _{\gamma }
|\det \Lambda (\beta ,\gamma )|^2,
\label{klevelh}
\end{eqnarray}
where we use the relation $\rho _A=\Lambda \Lambda ^{\dagger }$,
and the notation $\det \Lambda (\beta ,\gamma )$ means
the determinant of submatrix $\Lambda $ with row and
column index sets $\beta $ and $\gamma $.
When the cardinality $k=2,3$, we recover the
previous results (\ref{intuitive},\ref{3levelh}).
So, we do not need to calculate the eigenvalues of the
reduced density operator to find the concurrence hierarchy,
we can calculate the concurrence hierarchy directly
by summing the determinants of all $k$-by-$k$ submatrices
of $\Lambda $.

Next, we show the concurrence hierarchy cannot increase
under LOCC. We use the theorem proposed by Nielsen by
majorization scheme \cite{N}. 
For convenience, we use the same notations as that of Ref.\cite{B}
and Nielsen. The elements of vectors $x=\{ x_0^{\downarrow },
\cdots, x_{d-1}^{\downarrow } \} $
and $y=\{ y_0^{\downarrow },
\cdots, y_{d-1}^{\downarrow } \} $ are ordered in decreasing order.
We say that $x$ is majorized by $y$, $x\prec y$, if
$\sum _{j=0}^kx_j^{\downarrow }\le \sum _{j=0}^ky_j^{\downarrow },
k=0, \cdots, d-1$ and the equality holds when $k=d-1$.

Theorem 1 by Nielsen \cite{N}:
{\it $|\Psi \rangle $ transforms to $|\Phi \rangle $ using LOCC if and only
if $\lambda _{\Psi }$ is majorized by $\lambda _{\Phi }$,
\begin{eqnarray}
|\Psi \rangle \rightarrow |\Phi \rangle ~~~{\rm iff}~~~
\lambda _{\Psi }\prec \lambda _{\Phi }.
\end{eqnarray}
}

Now we propose our theorem by directly using Nielsen theorem.

Theorem 2: {\it $|\Psi \rangle $ transforms to $|\Phi \rangle $ using LOCC,
the concurrence hierarchy of $|\Psi \rangle $ is no less than 
that of $|\Phi \rangle $. And explicitly, if
$
|\Psi \rangle \rightarrow |\Phi \rangle $, then
$C_k(\Psi )\ge C_k(\Phi ) , ~k=1, \cdots, d$.
}

The proof of this theorem is as follows. Because of Nielsen theorem,
$|\Psi \rangle \rightarrow |\Phi \rangle $ then we have
$\lambda _{\Psi }\prec \lambda _{\Phi }$.
Because $-C_k, ~k=1, \cdots, d$ are isotone functions \cite{B},
i.e. if $\lambda _{\Psi }\prec \lambda _{\Phi }$ then
$-C_k(\Psi )\le -C_k(\Phi )$. Thus
we have $C_k(\Psi )\ge C_k(\Phi ) , ~k=1, \cdots, d$.
Here we mainly use the fact that each
$C_k$ are Schur-concave functions, see \cite{B}.

It is well known
that minus entropy function is isotone, so the
entanglement of formation cannot increase under LOCC.
Here we show the concurrence hirerarchy cannot increase under LOCC.

\section{Applications of concurrence hierarchy}
According to the theorem 2, if some of the relations
$C_k(\Psi )\ge C_k(\Phi ) , ~k=1, \cdots, d$ do not hold,
$|\Psi \rangle $ and $|\Phi \rangle $ can not be transformed to
each other by LOCC.
Here we analyze an example raised by Nielsen \cite{N},
\begin{eqnarray}
|\Psi \rangle =\sqrt{0.5}|00\rangle +\sqrt{0.4}|11\rangle
+\sqrt{0.1}|22 \rangle ,
\nonumber \\
|\Phi \rangle =\sqrt{0.6}|00\rangle +\sqrt{0.2}|11\rangle
+\sqrt{0.2}|22 \rangle .
\label{example3}
\end{eqnarray}
According to Nielsen theorem, neither
$|\Psi \rangle \rightarrow |\Phi \rangle $ nor
$|\Phi \rangle \rightarrow |\Psi \rangle $.
Here we analyze this example by calculating their concurrence
hierarchy. We can find
\begin{eqnarray}
C_2(\Psi )=0.29>C_2(\Phi )=0.28,
\label{2level}
\\
C_3(\Psi )=0.020<C_3(\Phi )=0.024.
\label{3level}
\end{eqnarray}
It follows from theorem 2 that neither
$|\Psi \rangle \rightarrow |\Phi \rangle $ nor
$|\Phi \rangle \rightarrow |\Psi \rangle $.
We can roughly interprete the reason as that the 2-level
entanglement of $|\Psi \rangle $ is larger than that of
$|\Phi \rangle $ (\ref{2level}),
but the 3-level
entanglement of $|\Psi \rangle $ is less than that of
$|\Phi \rangle $ (\ref{3level}). So we cannot transform them to
each other by LOCC.

It should be noted that the inverse of theorem 2 is not ture.
That means even 
we have $C_k(\Psi )\ge
C_k(\Phi ) , ~k=1, \cdots, d$, we are not sure
$|\Psi \rangle \rightarrow |\Phi \rangle $. Here we give an
example
\begin{eqnarray}
|\Phi '\rangle =\sqrt{0.5}|00\rangle +\sqrt{0.4}|11\rangle
+\sqrt{0.1}|22 \rangle ,
\nonumber \\
|\Psi '\rangle =\sqrt{0.55}|00\rangle +\sqrt{0.3}|11\rangle
+\sqrt{0.15}|22 \rangle .
\label{example4}
\end{eqnarray}
One can find the following relations
\begin{eqnarray}
C_2(\Psi ')=0.2925>C_2(\Phi ')=0.29,
\\
C_3(\Psi ')=0.02475>C_3(\Phi ')=0.020.
\end{eqnarray}
According to Nielsen theorem 
neither $|\Psi '\rangle \rightarrow |\Phi '\rangle $ nor
$|\Phi '\rangle \rightarrow |\Psi '\rangle $.
That means the concurrence hierarchy is not complete.
In the sense of classification pure bipartite states
by LOCC, Nielsen theorem is more powerful.
However, our result is mainly to quantify the entanglement by
concurrence hierarchy.

\section{Summuary and discussions}
The drawback of the concurrence hierarchy is that it is not complete
though the hierarchy consists of $d-1$ independent invariants.
We shoud note that
Vidal\cite{V1}, Jonathan and Plenio\cite{JP} and Hardy\cite{H0}
found a complete set of entanglement measures consists of
$d-1$ independent entanglement monotones.
In concurrence hierarchy, each
level of concurrence involves all parameters of a given pure state.
So, we can say that each concurrence in the hierarchy describes the
entanglement {\it globally}. For example, $C_2(\Phi )$ describes
all 2 level entanglement in a pure state $|\Phi \rangle $.
If two eigenvalues between $\lambda _{\Phi }^{\downarrow }$
and $\lambda _{\Phi}^{\downarrow }$ are different,
the concurrences in the hierarchy generally will be different.

In summary, we give the definition of concurrence hierarchy.
And we propose to use the concurrence hierarchy as measures
of entanglement. All concurrences in the hierarchy are zeros
for separable states except the
normalizaiton one. The concurrence hierarchy is invariant
under local unitary transformations. The concurrence
hierarchy cannot increase by using LOCC.
A simple and direct formula (\ref{klevelh}) is obtained
for concurrence hierarchy.
We also analyze some interesting examples by using
concurrence hierarchy.

Our result in this paper is a small step toward completely quantifying
the entanglement.
And we find
some interesting applications of concurrence hierarchy.
There are a lot of works need to be done along the direction of
this paper.
We just consider the case of
pure states. To study the concurrence hierarchy for mixed state
is difficult presently. Because even the first
non-trivial concurrence of a general mixed state
in d-dimension has not been obtained.
We even do not have a widely accepted operational way to
find whether a state is entangled. However, our result has
potential applications for mixed states. In particular, we give
the definition of concurrence hierarchy (\ref{hierarchy}),
it could shed light on how we should formulate them for mixed states.
We should note that the definition of concurrence hierarchy
(\ref{hierarchy}) is just for pure state.
To calculate the concurrence hierarchy for
mixed states, we need some formulas like the form of Wootters
in 2-dimension (\ref{magic}). Because we can not  
characterize separability only by the eigenvalues of density
matrix and reduced density matrices \cite{NK}.

As we already mentioned, even in classification of
pure states by LOCC, the theorem 2 is weaker than Nielsen theorem
though it has interesting applications.
But we actually raise an interesting question,
both $|\Psi \rangle $ and $|\Phi \rangle $ in (\ref{example3})
and 
$|\Psi '\rangle $ and $|\Phi '\rangle $ in (\ref{example4})
are incomparable by Nielsen theorem.
However, by using concurrence hierarchy, we show 
case (\ref{example3}) and case (\ref{example4}) are
belong to different groups. Then what's the essential
differences between the case (\ref{example3}) and the case
(\ref{example4})?

It is also interesting to consider other series of quantities to
quantify entanglement, for example, we can use
invariants $I_k=Tr(\Lambda \Lambda ^{\dagger })^{k+1}$ as
measures of entanglement. And quantum R\'enyi entropies
defined as $S_j={1\over {1-j}}log_2Tr(\Lambda \Lambda ^{\dagger })^j$,
see for example \cite{ZB,HW0}, also provide measures of entanglement.
Hopefully, quantum R\'enyi entropies can constitute a complete set
of measures of entanglement. And these measures of
entanglement work very well for the examples appeared in this
paper, i.e., they can determine whether a pure state can
be transformed to another by LOCC. However, a proof of
whether quantum R\'enyi entropies is complete or not is
necessary. It's also interesting to study whether we
can use concurrence hierarchy to study the mixed states,
the result in Ref.\cite{FJ} may be useful to this problem.
Some results about invariants for multipartite states
are already available\cite{Barnum}, it is worth to study
the corresponding concurrence hierarchy for multipartite states.

{\it Acknowlegements}: We would like
to thank J.Gruska, M.Hamada,
W.Y.Hwang, T.Shimono, X.B.Wang, A.Winter and H.Yura
for numerous useful discussions, and stimulating talks
in ERATO internal seminars. We also thank
M.M.Sinolecka, K.Zyczkowski, M.Kus for kind communications.

\end{document}